\documentclass[eqsecnum,aps,pra,twocolumn,groupedaddress,showpacs,showkeys]{revtex4}

\usepackage{graphicx}
\begin{document}

\newcommand{\be}{\begin{equation}}
\newcommand{\ee}{\end{equation}}

\title{The gravity of magnetic stresses and energy}

\author{Giuseppe Bimonte, Enrico Calloni  and Luigi Rosa}

\affiliation{Dipartimento di Scienze Fisiche, Universit\`{a} di
Napoli Federico II, Complesso Universitario MSA, Via Cintia
I-80126 Napoli, Italy;\\ INFN, Sezione di Napoli, Napoli, ITALY\\
}

\date{\today}

\begin{abstract}

In the framework of designing laboratory tests of relativistic
gravity, we investigate the gravitational field produced by the
magnetic field of a solenoid. Observing this field might provide a
mean of testing whether stresses gravitate as predicted by
Einstein's theory. A previous study of this problem by Braginsky,
Caves and Thorne predicted  that the contribution to the
gravitational field resulting from the stresses of the magnetic
field and of the solenoid walls  would cancel the gravitational
field produced by the mass-energy   of the magnetic field,
resulting in a null magnetically-generated gravitational force
 outside the solenoid. They
claim that this null result, once proved experimentally,  would
demonstrate the stress contribution to gravity. We show that this
result is incorrect, as it arises from an incomplete analysis of
the stresses, which neglects the axial stresses in the walls. Once
the stresses are properly evaluated, we find that the
gravitational field outside a long solenoid is in fact independent
of Maxwell and material stresses, and it coincides with the
newtonian field produced by the linear mass distribution
equivalent to the density of magnetic energy stored in a unit
length of the solenoid.   We argue that the gravity of Maxwell
stress can be directly measured  in the vacuum region inside   the
solenoid, where the newtonian noise is absent in principle, and
the gravity generated by Maxwell stresses is not screened by the
negative gravity of magnetic-induced stresses in the solenoid
walls.

\end{abstract}


\maketitle

\section{Introduction}

According to General Relativity, material and field stresses are
sources of gravity, because the active gravitational mass density,
in the relativistic analogue of Poisson's equation, is
proportional to $\rho+T$, where $\rho$ is the density of energy
and $T$ is the trace of the stresses \cite{MTW}. Stress-generated
gravity is very important in a number of problems. For example, in
astrophysics it affects the maximum mass of neutron stars, but if
one intends it, in a broad sense, as the gravity produced by the
spatial components of the momentum-energy tensor, it displays its
full power in cosmology, where it  may well be responsible of the
recently discovered accelerated expansion  of the Universe
\cite{glanz}.

As of today, there exists no direct experimental proof that
stresses indeed gravitate, and it  is clearly  of great interest
to investigate the possibility of  a {\it laboratory} experiment
to test this prediction of General Relativity. Unfortunately, this
is very difficult because in ordinary material bodies, of a size
that can be handled in a laboratory, the  trace of stresses is
many orders of magnitude smaller than the energy density
associated with the   mass density of the body, and therefore its
effects are negligible. However, it was realized thirty years ago
\cite{thorne} that a possible way to circumvent this difficulty is
by  observing the gravity of magnetic fields, which one expects to
exist because in General Relativity   all forms of energy (and
stresses) are sources of gravity. Magnetic fields are interesting
in this respect because, according to Maxwell theory, the energy
density of a magnetic field  has the same magnitude as the trace
of the Maxwell stress tensor and therefore this type of experiment
may provide an excellent tool to probe the gravity of stresses.
With this purpose, the authors of Ref. \cite{thorne} considered a
simple setup, in which the gravity produced by the magnetic field
of a long solenoid would be measured by means of a torsion
balance, having one of its test masses  near the solenoid. Of
course, the difficulty of the experiment is due to the fact that
magnetically-generated gravity is very weak, for experimentally
attainable magnetic fields. To get an estimate of the required
magnetic fields and balance sensitivity, one may temporarily
neglect all stresses and assume, on the basis of the equivalence
between mass and energy, that the  magnetically-generated
gravitational field near a long solenoid  is the same as that of a
cylindrical road, with a  linear mass density equal to the
magnetic energy (divided by the square of the speed of light $c$)
stored in a unit length of the solenoid. Even for very strong
magnetic fields,  the effect is very small, if one considers that
the mass density equivalent to the energy density of a magnetic
field of $10^5$ G is as small as $4.4 \times 10^{-13}$ ${\rm g\;
cm}^{-3}$. However, it was argued in \cite{thorne} that the
demands of the experiment could have soon be met, imagining
realistic improvements of the technology available in the
seventies, in cryogenic low-noise torque-balances and
superconducting solenoids.

When considering the  effect of stresses, one notices that two
types of stresses  may contribute to the gravitational field of
the solenoid: Maxwell stresses of the magnetic field and material
stresses that build up in the walls of the solenoid in response to
the applied magnetic field. In  Ref \cite{thorne} it was correctly
stated that the walls of the solenoid can be considered to be in
instantaneous mechanical equilibrium, because in the considered
setup the modulation frequency of the magnetic field is extremely
low (around $10^{-3}$ Hertz, which represents the typical
resonance frequency of a torque balance). The   conclusion drawn
in \cite{thorne} was that the inclusion of stresses would lead to
a null  magnetically-generated gravitational force (apart from the
newtonian noise caused by the stress-induced modulation of the
mass-density of the solenoid walls), because of a purported
cancellation occurring between the gravity of stresses and the
gravity of magnetic energy.

This result   appears  suspicious, from the point of view of  a
well-known   paradox, that was pointed out long ago by Tolman
\cite{tolman} in his investigations on the role of stresses as
source of gravity. Tolman found the paradox  while considering the
gravitational field of a static spherical impermeable box filled
with a fluid, which undergoes a spherically symmetric
    transformation that conserves the total energy, but causes a change of
pressure, like matter and antimatter annihilating into radiation.
One may think that, since the total energy of the system is
preserved, the change in pressure determines a change in the
active gravitational mass of the box, and a consequent change in
the gravitational field   outside the box. However, this inference
is in contradiction with  Birkhoff's theorem, which states that
the external gravitational field of a spherically symmetric body
is static and  therefore it is insensitive to whatever spherically
symmetric transformations may occur inside the box.   The Tolman
paradox was investigated in \cite{put}, where the crucial role of
the walls that keep the fluid confined was realized. It was shown
there that
  the stresses that  build up in the
walls in response to the transformation, give a negative
contribution to the active gravitational mass of the system, that
just compensates the pressure contribution from the fluid inside,
resulting in an overall unchanged total gravitational mass across
the transformation, as expected from Birkhoff's theorem. The same
problem has been   investigated again in a recent paper
\cite{ehlers}, leading to analogous conclusions (The key role of
the stresses in the  walls bounding a relativistic gravitating
systems has been discussed by us very recently, in connection with
the problem of determining the weight of a Casimir apparatus in a
weak gravitational field \cite{bimonte}).  The general lesson that
one learns from these studies is that the   gravitational field
{\it outside} a spherical body is independent of the stresses in
its interior, and it is determined solely by the mass-energy
content of the body. Since there is no reason to imagine that this
is true only for the spherical case, one  is led to suspect that
the results of \cite{thorne} may not be correct. This motivated us
to reconsider in detail the analysis of \cite{thorne}, and we
present here our findings. We realized that the null result found
in \cite{thorne}  was determined by a mistaken evaluation of the
stresses that build up inside the solenoid walls when the magnetic
field is present. In particular, the authors overlooked the axial
stresses that  arise in response to the axial electrodynamic
compression of the solenoid. Besides leading to an incorrect
result for the magnetically-generated gravitational field outside
the solenoid, this error led  the authors to overlook the large
newtonian noise originating from magnetic-induced changes in the
length of the solenoid.

After stresses are properly accounted for, our analysis shows, in
a general way, that the total magnetically-generated gravitational
mass, measured far from the solenoid, is independent of the
stresses and is just equal to the total magnetic energy (divided
by the square of the speed of light $c^2$), in accordance with
one's intuition and in agreement with earlier studies on the
Tolman paradox. We then consider the field near a long solenoid,
and we show   that the magnetically-generated gravitational field
is different from zero, and as expected it is equivalent to the
newtonian field generated by a linear mass-density that is equal
to the instantaneous magnetic energy per unit length (divided by
$c^2$) stored in the solenoid. Since the near field outside the
solenoid, like the far field, is independent of the stresses, we
conclude that observation of the external field cannot be used  to
test the gravity of stresses. Moreover, measuring this
magnetically-generated field  will be very hard, because we
estimate that magnetic-induced changes in the length of the
solenoid  produce a newtonian noise that is many order of
magnitudes larger than the magnetically generated gravity. This by
no means implies, however, that the gravity of stresses is not
observable in this setup, because in the vacuum region {\it
inside} the solenoid  the gravity produced by Maxwell stresses is
not screened by the negative gravity  of the material stresses in
the walls, and therefore it contributes to the field as much as
the density of magnetic energy. Moreover, it is expected that the
newtonian noise will be much less of a problem, because in the
ideal case of a infinitely long and perfectly axially symmetric
solenoid, newtonian noise inside the solenoid is strictly zero.

The paper is organized as follows: in Sec. 2  we  derive, within
Linearized Theory for   General Relativity, the
magnetically-generated gravitational pull exerted on a test
particle by a solenoid carrying a quasi-static magnetic field. In
Sec. 3 we analyze in detail the contributions from Maxwell and
material stresses and we prove that outside the solenoid they
cancel each other. Sec. 4  deals with the problem of newtonian
noise, while Sec. 5 contains a discussion of the results and our
conclusions. Finally, in the Appendix we provide explicit formulae
for the material stresses that build up within the walls of an
idealized solenoid.

\section{The gravity  of a quasi-static magnetic field}

In this Section we   estimate the pull $F_i$ exerted on a test
particle at rest,  by the magnetically-generated  gravitational
field of a solenoid ${\cal S}$,  producing a quasi-static magnetic
field $B$. Since the gravitational fields involved are extremely
small, non-linear effects are negligible and we can safely study
the problem using the simple Linearized Theory  for Einstein's
General Relativity \cite{MTW}. In this approximation, the
gravitational field $g_{\mu \nu}$ is written as \footnote{in what
follows, Greek letters denote space-time indices, while Latin
letters denote space indices.}: \be g_{\mu \nu}=\eta_{\mu
\nu}+h_{\mu \nu}\;,\ee where $\eta_{\mu \nu}={\rm
diag}\{-c^2,1,1,1\}$ is the flat Minkowski metric, and $h_{\mu
\nu}$ represents a weak gravitational field. We further split
$h_{\mu \nu}$ as \be h_{\mu \nu}= h_{\mu
\nu}\vert_{B=0}+\gamma_{\mu \nu}\;,\ee where $h_{\mu
\nu}\vert_{B=0}$ is the field that exists when the solenoid is
turned off, while $\gamma_{\mu \nu}$ is the magnetically-generated
field that is present when the magnetic field $B$ is turned on.
The field $h_{\mu \nu}\vert_{B=0}$ includes the background
gravitational existing in the laboratory, together with the small
field generated by the walls of the solenoid when no currents flow
in it.

To linear order, the   pull $F_i$ on a test particle of mass $m$
arising from the magnetically-generated gravitational field is:
\be F_i =-m \,(\Gamma^i_{00}-\Gamma^i_{00}\vert_{B=0})=
\frac{1}{2}\,m \,
\partial_i \,\gamma_{00}\;,\ee where $\Gamma^i_{00}$ are Christoffel symbols.
 For a quasi-static magnetic field, Linearized Theory
gives the following Equations for $\gamma_{\mu \nu}$: \be
\triangle \bar{\gamma}_{\mu \nu}=- \frac{16 \pi\,G}{c^4}\,T_{\mu
\nu}\;.\label{linear}\ee In these Equations, $\triangle$ denotes
the flat space-time laplacian, and $\bar {\gamma}_{\mu \nu}$ is
the field: \be \bar{\gamma}_{\mu \nu}= \gamma_{\mu \nu}-
\frac{1}{2}\eta_{\mu \nu} \gamma,\ee where $\gamma=\eta^{\mu \nu}
\gamma_{\mu \nu}$. The above equations have to be supplemented by
the Lorenz gauge conditions, which for a static field imply: \be
\partial_i\, \bar{\gamma}^i_{\mu}=0\;.\label{gauge}\ee It is important to bear in mind
that, according to the definition of $\gamma_{\mu \nu}$, the
energy-momentum tensor $T_{\mu \nu}$ appearing on the r.h.s. of
Eqs. (\ref{linear}) represents the sole contribution to the total
energy-momentum tensor that arises when the magnetic field is
turned on. The solenoid being at rest, and the magnetic field
being quasi-static, the non-vanishing components of $T^{\mu \nu}$
read: \be T^{00}=\delta \rho_{\rm walls} +  {\cal E}_{\rm
mag}/c^2\;,\ee \be T^{ij}=T^{ij}_{\rm walls}+T^{ij}_{\rm
mag}\;.\label{spliTij}\ee In the above equations, $\delta
\rho_{\rm walls}$ represents the change in the (classical)
mass-density of the solenoid walls resulting from possible
deformations of the solenoid determined by the magnetic field
\footnote{we neglect here the tiny relativistic correction to the
mass density of the walls resulting from the change of the walls
internal-energy, as determined by their deformation.}, while
$T^{ij}_{\rm walls}$ denote the extra mechanical stresses that
build up within the solenoid walls when the field is turned on.
Note that $T^{ij}_{\rm walls}$ does not include the mechanical
stresses resulting from the weight of the solenoid  and from the
external forces  exerted on the solenoid walls by the mounts that
hold it. Finally, ${\cal E}_{\rm mag}=B^2/(8\,\pi)$ denotes the
density of magnetic energy, while $T^{ij}_{\rm mag}$ is the
Maxwell tensor: \be T_{\rm mag}^{ij}= \frac{1}{4
\pi}\left(\frac{1}{2}B^2\,\delta^{ij} \,-B^i\,B^j
\right)\;.\label{Tijmag1}\ee Upon solving Eqs. (\ref{linear}) it
is easy to obtain for the pull $F_i$ the following expression: \be
F_i =-m \, \partial_i\,\left(\delta \Phi_{\rm walls}+  \psi
\right),\label{varforPN}\ee where \be \delta \Phi_{\rm walls} =-G
\int d^3 {\bf y} \frac{\delta \rho_{\rm walls}}{|{\bf x}-{\bf
y}|}\;, \label{newpot}\ee and \be \psi =-G \int \frac{d^3 {\bf
y}}{|{\bf x}-{\bf y}|}\left( \frac{{\cal E}_{\rm mag}+ T^{ii}_{\rm
walls}+ T^{ii}_{\rm mag}}{c^2}\right) \;. \label{varpsif}\ee Of
the two terms appearing on the r.h.s. of Eq. (\ref{varforPN}),
that involving $\delta \Phi_{\rm walls}$ just represents a purely
classical "newtonian noise", and  we postpone to Sec.5 a
discussion of its consequences.  The interesting term for us   is
the contribution proportional to $\psi$, that represents the
magnetically-generated gravitational field.   We see that  $ \psi$
coincides with the classical gravitational field generated by and
effective mass distribution $\rho_{\rm eff}$ equal to: \be
\rho_{\rm eff}\,=\,\frac{1}{c^2}\left( {\cal E}_{\rm mag} +
 {T^{ii}_{\rm walls}}
 +T^{ii}_{\rm mag} \right)\;.\label{rhoeff}\ee  This is
a rather complicated formula, for it involves the trace of the
stresses $T_{\rm walls}^{ij}$  in the solenoid walls. It is
convenient to define the total "effective gravitational mass"
$M_{\rm eff}$, as  the integral over all space of $\rho_{\rm
eff}$: \be M_{\rm eff}=\int_{\rm All\; space}\!\!\!\!\! d^3
x\,\rho_{\rm eff}\,.\ee We split $M_{\rm eff}$ as: \be M_{\rm
eff}=M_{\rm mag \,en}+M_{\rm str}\;,\label{splitmeff}\ee where
$M_{\rm mag\,en}$ is the mass associated with the total magnetic
energy $E_{\rm mag}$ \be M_{\rm mag\,en} =\frac{1}{c^2}\,\int_{\rm
All\; space} \!\!\!\!\!d^3 {\bf x}\,{\cal E}_{\rm mag}\equiv
\frac{E_{\rm mag}}{c^2}\;,\ee  while
  $M_{\rm str}$ is associated with
Maxwell and material stresses: \be M_{\rm str} \equiv
\frac{1}{c^2}\,\int_{\rm All\; space} \!\!\!\!\!\!\!\!\!d^3 {\bf
x}\;( T^{ii}_{\rm walls}+T^{ii}_{\rm mag})\,.\label{totmef}\ee
Note that both the integrals for $M_{\rm mag \,en}$ and $M_{\rm
str}$ exist, because at large distances $R$ from the solenoid, the
magnetic field falls off like $R^{-3}$ and then ${\cal E}_{\rm
mag}$ and $T^{ij}_{\rm mag}$ both decay as $R^{-6}$. The existence
of a contribution  to $M_{\rm eff}$, such as $M_{\rm mag \,en}$,
arising from the magnetic energy is not surprising in view of the
equivalence between energy and mass, established in the Theory of
Special Relativity. On the contrary, the contribution $M_{\rm
str}$ from the stresses represents a true General Relativistic
effect. In the next Section it will be proven that $M_{\rm str}$
is always zero, at mechanical equilibrium.

\section{The contribution from stresses}

To be definite, we imagine that the solenoid ${\cal S}$ is hanging
by a suitable set of threads, and that apart from the suspension
points its surface is free.    Now,  upon taking the spatial
divergence of both sides of Eq. (\ref{linear}), and then using the
gauge condition Eq. (\ref{gauge}), we obtain \be
\partial_i \, (T^{ij}_{\rm walls}+
  \,T^{ij}_{\rm mag}) =0\,. \label{equil}\ee
Outside the solenoid walls, where $T^{ij}_{\rm walls}=0$, the
above equations are satisfied as a consequence of the static
Maxwell Equations in vacuum: \be {\bf \nabla \cdot B}={\bf
0}\;,\;\;\;\;{\bf \nabla \times B}={\bf 0}\;.\ee Inside the
solenoid walls, instead, Eqs. (\ref{equil}) express  the local
balance between  electrodynamic forces and material stresses, at
mechanical equilibrium. At points on the boundary $\partial {\cal
S}$ of the solenoid walls, Eqs. (\ref{equil}) must be supplemented
by the following boundary condition  \be n_i\,(T^{ij}_{\rm
walls}+T^{ij}_{\rm mag})\vert_{\rm ins}\,=n_i\,T^{ij}_{\rm
mag}\vert_{\rm out}\;, \label{bc}\ee where $n^i$ is the normal to
the surface of the solenoid walls, oriented outwards the solenoid,
and the suffixes ${\rm ins}$  (${\rm out}$) denote the values of
the fields immediately inside (outside) the solenoid walls. Eq.
(\ref{bc}) expresses the fact that the total electrodynamic
self-force on the solenoid is zero, and therefore the  threads
that support it do not apply any extra force when the magnetic
field is turned on. Using Eq. (\ref{equil}) and the boundary
condition Eq. (\ref{bc}), we can now show that $M_{\rm str}$ is
always zero. For this purpose, we note that at all points not
lying on the boundary $\partial {\cal S}$ of ${\cal S}$, Eq.
(\ref{equil}) implies the identity: \be ( T^{ii}_{\rm
walls}+T^{ii}_{\rm mag})=\partial_j\,[( T^{ij}_{\rm
walls}+T^{ij}_{\rm mag})\, x^i]\;.\ee Upon substituting this
expression for $T^{ii}_{\rm walls}+T^{ii}_{\rm mag}$ into the
r.h.s. of Eq. (\ref{totmef}), and then performing the integral of
the total divergence by Gauss theorem, we obtain for $M_{\rm str}$
the expression: $$ M_{\rm str}= \int_{\rm
\partial {\cal S}} \!\!d^2 \sigma \,x^j\,
n^i [(T^{ij}_{\rm walls}+T^{ij}_{\rm mag})\vert_{\rm
ins}-T^{ij}_{\rm mag}\vert_{\rm out}] \,+$$ \be \;\;\;\;\;+\lim_{R
\rightarrow \infty}\int_{S_R} \!\!\!\!\!\!d^2 \sigma \,
T^{ij}_{\rm mag} \,x^j\, n^i\;,\ee where $S_R$ denotes a
two-sphere of radius $R$ centered at any point inside the
solenoid. Now, the first integral on the r.h.s. is zero because of
the boundary condition Eq. (\ref{bc}), and the second vanishes
because $T^{ij}_{\rm mag}n^i x^j$ falls off as $R^{-5}$.
Therefore, as promised, we obtain \be M_{\rm
str}=0\;.\label{mstr}\ee The conclusion is that, independently of
the shape of the solenoid and of the detailed distribution of the
stresses inside its walls, the general conditions of mechanical
equilibrium as encoded in Eqs. (\ref{equil}) and Eqs. (\ref{bc})
imply that the combined contribution of Maxwell and material
stresses to the {\it total} gravitational mass of the solenoid
vanishes. Therefore, the total effective gravitational mass
associated with the magnetic field is equal to $M_{\rm mag \,en}$:
\be M_{\rm eff}=M_{\rm mag \,en} \label{magen}\;.\ee  The
gravitational field that is  observed far from the solenoid when
the magnetic field is turned on is then equal to that of a point
charge with mass $M_{\rm mag \,en}$, placed at the position of the
solenoid.

Obviously, Eq. (\ref{mstr}) does not imply that the magnetically
generated stresses produce no gravity at all, because it only
states that Maxwell and material stresses cancel each other {\it
on average}, namely after integrating over all space. While this
is sufficient to  conclude that stresses do not contribute to the
far-field, it still remains the possibility that stresses produce
significant gravitational effects in the vicinity of  the
solenoid, for the near field probes also the detailed {\it
spatial} distribution of the stresses.   The study of the near
field is clearly much more complicated in general,  because it
requires a detailed determination of the mechanical stresses
inside the walls of the solenoid. The study of the stresses that
arise in a solenoid generating a strong magnetic field has
received much attention in the literature over the years, in view
of its great practical importance (see for example Ref.\cite{melv}
and References therein), and   in general it is a difficult
problem, that involves making a definite model for the
constitutive equations characterizing the material, and it usually
requires numerical tools. We shall not discuss this difficult
problem here, and we  content ourselves with a few simple
considerations that can be drawn  on the basis of general
mechanical  equations, without any consideration of  specific
constitutive equations. To simplify the problem,  we consider
below a very long cylindrical solenoid and we discuss separately
the gravitational field outside and inside the solenoid.

\subsection{The external near field}

We consider, as in Ref.\cite{thorne}, a very long cylindrical
solenoid, constituted by a (non magnetic) pipe with inner and
outer radii $R_1$ and $R_2$ respectively, and length  $L \gg R_2$.
We suppose for simplicity that the electric current producing the
magnetic field flows along the inner surface of the pipe, in the
positive azimuthal direction, and that it has a uniform surface
density $j$. We let $\{x,y,z\}$ a cartesian coordinate system
whose $z$ axis coincides with the solenoid axis, and whose origin
lies at the center of the solenoid, and we let $r=\sqrt{x^2+y^2}$
the distance from the solenoid axis.

Axial symmetry obviously implies that the effective mass density
$\rho_{\rm eff}$ in Eq. (\ref{rhoeff}) is a function only of $r$
and $z$. As a first step, we show that   $\rho_{\rm eff}$ is
significantly different from zero only inside the solenoid, i.e.
for $r \le R_2$ and $|z| \le L/2$. This is obvious for the
contribution to $\rho_{\rm eff}$ arising from the material
stresses, because $T^{ij}_{\rm walls}$ vanish outside the solenoid
walls.   Then, upon noting that \be T^{ii}_{\rm mag}={\cal E}_{\rm
mag}\;,\label{traceB}\ee as can be seen by taking the trace of the
Maxwell stresses in Eq. (\ref{Tijmag1}), we see that the
contribution to $\rho_{\rm eff}$ arising from the magnetic field
is equal to twice ${\cal E}_{\rm mag}/c^2$. We can   estimate the
integral of ${\cal E}_{\rm mag}$ outside the solenoid as follows:
the external magnetic field coincides with the   field of a
cylindrical magnet having length $L$ and  radius $R_1$,  carrying
a uniform magnetization $m=j/c$ along the positive $z$-direction.
The field of such a magnet coincides with the sum of the fields
${\bf B}_1$ and ${\bf B}_2$ produced by  the opposite surface
distributions of magnetic  charges on the opposite  caps of the
magnet (at $z=\pm L/2$), with uniform surface densities
$\sigma_m=\pm j/c$. The total energy $E^{\rm ext}_{\rm mag}$ of
the external field can then be estimated to be \be E^{\rm
ext}_{\rm mag}=\frac{1}{8 \pi}\int_{\rm outside}\!\!\!\!\!\!\! d^3
x\,(B^2_1+B_2^2)+\frac{1}{4 \pi}\int_{\rm outside} \!\!\!\!\!\!\!
d^3 x\,{\bf B}_1 \cdot {\bf B}_2\,.\label{enout}\ee The first
integral on the r.h.s. of the above Equation represents the sum of
the magnetic energies of two isolated pole distributions at $z=\pm
L/2$. Therefore,  it is  independent of the solenoid length $L$,
and on dimensional grounds one expects it to be of the form: \be
\frac{1}{8 \pi}\int_{\rm outside}\!\!\!\!\!\!\! d^3
x\,(B^2_1+B_2^2)= \frac{B^2_{\rm in}}{8 \pi}\,2\,A\,
R_1^3\;,\label{self}\ee where $B_{\rm in}=4 \pi j/c$ is the
magnetic field inside the solenoid, and $A$ is some numerical
constant. As for the second integral on the r.h.s of Eq.
(\ref{enout}), it  represents the interaction energy  among the
two poles of the magnet, and it can be approximated as the
interaction energy of two opposite point-like magnetic charges of
magnitude $q_m=\pi R_1^2 j/c$ at distance $L$: \be \frac{1}{4
\pi}\int_{\rm outside} \!\!\!\!\!\!\! d^3 x\,{\bf B}_1 \cdot {\bf
B}_2\simeq \frac{q_m^2}{L}=\frac{2\,\pi^2
\,j^2\,R_1^4}{c^2\,L}=\frac{B^2_{\rm
in}\,R_1^4}{8\,L}\,.\label{int}\ee Adding  up Eq. (\ref{self}) and
Eq. (\ref{int}), we obtain for $E_{\rm mag}^{\rm ext}$ the
expression: \be E_{\rm mag}^{\rm ext} \simeq \left(2
\,A+\frac{\pi\,R_1}{L}\right) \frac{B_{\rm in}^2}{8 \pi} \,
R_{1}^3\ee On the other hand, the internal magnetic energy $E^{\rm
int}_{\rm mag}$ can be estimated to be \be E^{\rm int}_{\rm
mag}=\frac{B_{\rm in}^2}{8 \pi} \times \pi R_1^2\,L ,\ee and
therefore we obtain for the ratio of $E^{\rm ext}_{\rm mag}/E^{\rm
int}_{\rm mag}$ the estimate: \be \frac{E^{\rm ext}_{\rm
mag}}{E^{\rm int}_{\rm mag}}=\frac{2 A}{\pi}\,\frac{R_1}{L}+
\left(\frac{R_1}{L}\right)^2,\ee which shows that $E^{\rm
ext}_{\rm mag}$ becomes negligible with respect to $E^{\rm
int}_{\rm mag}$ for $R_1/L \ll 1$.

Consider now a point $P$ in the vicinity of the solenoid, but far
from its ends. The above  estimation of the external magnetic
stresses and energy shows that the magnetically-generated
gravitational field at $P$ is determined by the stresses and the
magnetic energy that are present inside the solenoid and within
its material walls.  Since far from the solenoid's ends the
magnetic field and the material stresses are approximately
independent of the $z$ coordinate, we see from Eq. (\ref{varpsif})
that the field $ \psi$ at $P$ coincides with the classical field
of an infinite cylindrical rod, with a uniform linear mass density
$\sigma_{\rm eff}$   equal to: \be \sigma_{\rm
eff}\,=\,\frac{1}{c^2}\int_0^{R_2} dr\,2 \pi r\,\left( {\cal
E}_{\rm mag} +  {T^{ii}_{\rm walls}}
 +T^{ii}_{\rm mag} \right)\;.\label{sigmaeff}\ee
   Now, we can
split $\sigma_{\rm eff}$   analogously to what we did with $M_{\rm
eff}$ in Eq. (\ref{splitmeff}): \be \sigma_{\rm eff}=\sigma_{\rm
mag\;en}+\sigma_{\rm str}\;,\ee where \be \sigma_{\rm
mag\;en}=\,\frac{1}{c^2}\int_0^{R_2} dr\,2 \pi r\, {\cal E}_{\rm
mag}\equiv \frac{{\tilde {\cal E}}_{\rm mag}}{c^2}
\;,\label{sigmamagen}\ee with ${\tilde {\cal E}}_{\rm mag}$ the
magnetic energy per unit length of the solenoid, and \be
\sigma_{\rm str}=\,\frac{1}{c^2}\int_0^{R_2} dr\,2 \pi r\, \left(
 {T^{ii}_{\rm walls}}
 +T^{ii}_{\rm mag} \right)  \label{sigmastr}\;.\ee We can easily see that $\sigma_{str}$ vanishes.
 Indeed, neglecting the
  contributions to $M_{\rm str}$ from the external magnetic
field, which we have seen to be small, as well as the contribution
from the small region near the solenoid's ends, we can then
express $M_{\rm str}$ as \be M_{\rm str}= L \,\sigma_{\rm
str}\;.\ee Since, according to Eq. (\ref{mstr}) $M_{\rm str}$ is
zero, it follows at once \be \sigma_{\rm str}=
0\;.\label{sigstr}\ee We conclude that also   near the solenoid
the magnetically-generated gravitational field $\psi$ is
independent of the stresses, and it simply coincides with the
field generated by a cylindrical distribution of mass, having a
linear density that is equal to the instantaneous magnetic energy
stored in the solenoid (divided by $c^2$) per unit length: \be
\sigma_{\rm eff}\,=\frac{ {\tilde {\cal E}}_{\rm mag}}{c^2}
\;.\label{sigmaefffin}\ee

These results are in sharp contrast with the findings of
Ref.\cite{thorne}, where  it was  concluded   that the
contribution
 from Maxwell and material
stresses is {\it different from zero}, and of such a magnitude as
to cancel the gravitational field produced by the mass-energy of
the magnetic field, resulting in a null magnetically-generated
gravitational field $ \psi$ outside the solenoid. A detailed
analysis of the sketchy computations in \cite{thorne} shows that
this incorrect conclusion arose from an incomplete evaluation of
the material stresses that build up inside the solenoid walls, as
the authors only considered the effect of the radial
electrodynamic forces pushing to increase the radius of the
solenoid, but they overlooked the existence of an axial force
tending to compress the solenoid \cite{kapi}. When the
contribution from the axial stresses is accounted for, our result
Eq. (\ref{sigmaefffin}) is recovered. As a further check of the
fundamental Eq. (\ref{sigstr}), in the Appendix we provide the
explicit  formulae for the material stresses that build up within
the walls of an idealized solenoid.

\subsection{Internal field}

We consider now the gravitational field in the vacuum region in
the {\it interior} of the solenoid, i.e. for $r < R_1$. Since
$T^{ij}_{\rm walls}$ is zero for $r < R_1$, the field $\psi$
coincides with the classical potential generated by a linear mass
density $\sigma_{\rm eff}^{\rm int}(r)$: \be \sigma_{\rm eff}^{\rm
int}(r)\,=\,\frac{1}{c^2}\int_0^{r} dr'\,2 \pi r'\,\left( {\cal
E}_{\rm mag}
 +T^{ii}_{\rm mag} \right) .\label{sigmaint0}\ee
Differently from the external region, in the interior of the
solenoid Maxwell
 stresses are not screened by material stresses, and therefore they do contribute to
 the {\it internal} gravitational field.
Upon recalling that $T^{ii}_{\rm mag}={\cal E}_{\rm mag}$, see
Eq.(\ref{traceB}), we see that Maxwell stresses contribute to the
internal field as much as magnetic energy, and then we can rewrite
Eq. (\ref{sigmaint0}) as
 \be \sigma_{\rm
eff}^{\rm int}(r)\,= \frac{1}{c^2}\int_0^{r} dr'\,2 \pi r'\,2
\,{\cal E}_{\rm mag} \equiv \frac{2\,{\tilde {\cal E}}_{\rm
mag}^{\rm int}(r)}{c^2}\;.\label{sigmaint}\ee It is interesting to
consider a
  solenoid with thin walls. Since in such a case the magnetic energy contained
in the region of space occupied by the solenoid walls is
negligible, we have \be  {\tilde {\cal E}}_{\rm mag}^{\rm
int}(R_1) \simeq  {\tilde {\cal E}}_{\rm mag}\;,\ee and therefore
Eq. (\ref{sigmaint}) implies that a test mass placed immediately
inside the solenoid would feel an oscillating pull towards the
solenoid's axis that is {\it twice} as strong as the pull observed
just outside the solenoid: \be F(R_1) = 2\, F(R_2)\;.\ee This
result arises because, for $r < R_1$, the gravity originating from
Maxwell stresses (the second term inside the brackets in Eq.
(\ref{sigmaint0})) is not screened by the negative gravity of the
magnetic-induced stresses in the walls of the solenoid.

\section{The newtonian noise}

Producing strong magnetic fields and designing sensitive  torque
balances may not be enough  to ensure that one would be able to
actually observe the magnetically generated gravitational field $
\psi$. For that to be possible,  one has to make sure that the
newtonian noise $\delta \Phi_{\rm walls}$ is not exceedingly large
compared to $ \psi$. The order-of-magnitude estimate presented
below shows that there are little prospects of measuring $ \psi$
outside the solenoid, for  we estimate that outside the solenoid
$\delta \Phi_{\rm walls}$ is about nine order of magnitudes larger
that $ \psi$. At the end, we shall briefly comment on the chances
of measuring $ \psi$ {\it inside} the solenoid, where the
newtonian noise is expected to be much smaller.

 As we pointed out in the previous Sections,
$\delta \Phi_{\rm walls}$ comes about because electrodynamic
forces   deform    the solenoid walls, resulting in a change of
shape and density of the walls. An accurate determination of
$\delta \Phi_{\rm walls}$ requires a detailed model for the
solenoid, and is beyond the scope of the present paper. We shall
content ourselves with simple considerations based on
order-of-magnitude estimates.

We consider separately  radial electrodynamic forces, that tend to
increase the radius of the solenoid,  and axial electrodynamic
forces, that tend to make the solenoid shorter. Radial forces were
the only source of newtonian noise that was considered in
\cite{thorne}, because, as observed earlier, the   authors  did
not take account of the axial compression of the solenoid. In
principle, radial deformations are innocuous because, for a
perfectly cylindrical solenoid, a symmetric radial deformation
does not alter the axial mass-density of the solenoid, and
therefore it produces no newtonian noise. Real solenoids of course
are not perfectly symmetrical, and therefore one expects that
slightly asymmetrical radial deformations will actually produce
some noise. A possible remedy for this problem was pointed out in
Ref. \cite{thorne}, and consists in averaging over azimuthal
inhomogeneities in the radial deformation, by setting the solenoid
in rotation around its axis, with an angular frequency much larger
than the modulation frequency of the magnetic field.

As we shall now see, the real trouble comes from the axial
compression of the solenoid. To estimate the newtonian noise
introduced by this compression, we consider a cylindrical long
solenoid of length $L$, whose walls have a cross-sectional area
$A_{\rm walls}$.  We assume for simplicity that the  axial
compression $T^{zz}_{\rm walls}$ is uniform throughout the section
of the walls, and that it does not exceed the elastic limit of the
material. If we let $F_{\rm ax}$ the total axial compression \be
F_{\rm ax}=\int_{R_1}^{R_2} dr\,2 \pi r\, T^{zz}_{\rm walls}\;,\ee
from Hook's law we estimate that the length of solenoid will
suffer a fractional change of magnitude: \be
 \frac{\delta\, L}{L} =-\frac{1}{E} \times \frac{F_{\rm ax}}{A_{\rm walls}}\;,\label{hook0}\ee where
 $E$ is the Young modulus for the
material of walls.
In the Appendix we show that, sufficiently far from the
end-points, the axial compression $F_{\rm ax}$ has magnitude: \be
F_{\rm ax}= {\tilde {\cal E}}_{\rm mag}\:.\label{compr}\ee Using
this formula in the r.h.s. of Eq. (\ref{hook0}), we obtain an
estimate of the relative change in the solenoid length:\be
 \frac{\delta\, L}{L} = -\frac{1}{E} \times
 \frac{{\tilde {\cal E}}_{\rm mag}}{A_{\rm walls}}\;.\label{hook}\ee Consider now the total
mass $\sigma_{\rm walls}$  per unit length of the solenoid.
Obviously, under a change $\delta L$ in the solenoid length,
$\sigma_{\rm walls}$ changes  by the amount \be \delta\,
\sigma_{\rm walls}=-\frac{\delta\, L}{L}\, \sigma_{\rm
walls}\;.\ee Then, from Eq. (\ref{hook}) we obtain: \be \delta\,
\sigma_{\rm walls}=\frac{ {\tilde {\cal E}}_{\rm mag}}{E} \times
\frac{\sigma_{\rm walls}}{A_{\rm walls}}=\frac{{\tilde {\cal
E}}_{\rm mag}}{E}\,\rho_{\rm walls}\,\;,\ee where $\rho_{\rm
walls}=\sigma_{\rm walls}/A_{\rm walls}$ is the mass density of
the material for the walls. Having estimated the change $\delta
\sigma_{\rm walls}$ in the linear mass-density of the solenoid, we
can easily obtain an estimate for the ratio $  \psi/\delta
\Phi_{\rm walls}$ among the magnetically-generated field $\psi$
and the newtonian noise. Since the former is proportional to
$\sigma_{\rm eff}$ and the latter to $\delta\, \sigma_{\rm
walls}$, we find \be \frac{  \psi}{\delta \Phi_{\rm walls}}=
\frac{\sigma_{\rm eff}}{\delta \sigma_{\rm walls}}=\frac{ {\tilde
{\cal E}}_{\rm mag}}{c^2}\times \frac{E}{ {\tilde {\cal E}}_{\rm
mag}\,\rho_{\rm walls}}=\frac{E}{c^2\,\rho_{\rm walls}}\;,\ee
where in the  second passage we used Eq. (\ref{sigmaefffin}). It
should be noted that the  result is independent of the strength of
the magnetic field. In the case of stainless steel, which has $E=2
\times 10^{11}\;{\rm N}/{\rm m}^2$ and $\rho=8\,{\rm g}/{\rm
cm}^3$, we obtain: \be \frac{  \psi}{\delta \Phi_{\rm walls}}=2.8
\times 10^{-10}\;,\ee and we see that the newtonian noise is over
nine order magnitudes larger than the magnetically-generated
field.

This elementary analysis shows that it will be extremely difficult
to observe the oscillating field $ \psi$ outside the solenoid.
However, the newtonian noise should be much less of a problem {\it
inside} the solenoid, which we showed to be the interesting region
for the purpose of testing the gravity of stresses. This is so
because, in the ideal case of an infinitely long and perfectly
axially symmetric solenoid, the newtonian noise inside the
solenoid is strictly zero.

\section{Conclusions}

According to General Relativity, stresses act as a source of
gravity on the same footing as energy. While stress-generated
gravity is normally negligible, it is though to play an important
role in astrophysics, where it contributes to determining the
maximum mass of neutron stars, and it is perhaps determinant in
cosmology, where "negative" pressure-generated gravity may be the
cause of the recently discovered accelerated expansion of the
Universe. The importance of these problems makes it highly
desirable to design a laboratory test, still lacking as we write,
to verify if stresses actually gravitate as predicted by General
Relativity, or not. A test of this sort was proposed long ago in
\cite{thorne}, and it involved measuring the gravitational pull on
a test mass placed outside a long solenoid, carrying a slowly
alternating current. The   conclusion was that in General
Relativity the oscillating magnetic field inside the solenoid
produces a null gravitational force on the test mass, because the
attractive gravity generated by the energy and stresses of the
magnetic field was   found to cancel against the negative gravity
generated by the material stresses that build up inside the
solenoid walls. In this paper we demonstrated that this  result is
incorrect, as it hinges on a mistaken analysis of the material
stresses, in which the electrodynamic axial compression of the
solenoid was overlooked. After amending this mistake, we found
that the contribution to the external gravitational field from
Maxwell stresses and material stresses within the walls cancel
each other,  and therefore the resulting gravitational field is
determined solely by the linear density of magnetic energy stored
inside the solenoid. Thus observation of the   external field
cannot be used to test the gravity of stresses. Moreover,
observing this field is extremely unlikely because of the enormous
newtonian noise that results from small changes in the length of
the solenoid caused by the axial electrodynamic compression.

The interesting region for testing the gravity of stresses is the
one inside the solenoid, because there the gravity of Maxwell
stresses is not screened by the gravity of material stresses
existing in the solenoid walls, and therefore they contribute as
much as the magnetic energy  in generating gravity. In the
internal region the newtonian noise should also be much less of a
problem, because in the ideal case of a long solenoid, with
perfect axial symmetry, newtonian noise is zero. The major
experimental difficulty that we foresee, apart from control of the
residual noise resulting from asymmetries of the solenoid, is to
find means of accurately measuring the gravitational field in the
presence of strong magnetic fields.

\acknowledgments The authors would like to thank Prof. K.S. Thorne
for  valuable comments that helped greatly improving the
manuscript. The work of L. Rosa has been partially supported by
PRIN {\it FISICA ASTROPARTICELLARE}.

\section{Appendix}

In this Appendix we provide the explicit formulae for the stresses
that build up within the walls of the idealized solenoid
considered in Sec. 3.A, consisting of a cylindrical pipe carrying
a uniform azimuthal current concentrated on its inner face. The
expressions presented below provide an explicit verification of
the important formulae, Eq. (\ref{sigstr}) and Eq. (\ref{compr}).

We consider first the effect of the radial magnetic pressure $P$,
 on the inner face of the pipe. Far from the solenoid's
ends, $P$ is uniform and its magnitude is equal to the radial
component of the Maxwell stress tensor $T^{rr}_{\rm mag}$ inside
the solenoid: \be P=\frac{B_{\rm in}^2}{8 \pi}\;.\ee  This radial
pressure determines transverse stresses $T^{rr}_{\rm walls}(r)$
and $T^{\phi \phi}_{\rm walls}(r)$ in the pipe's walls, whose
expressions are well known \cite{young} and read:
$$ T^{rr}_{\rm walls}=P
\frac{R_1^2}{R_2^2-R_1^2}\left(\frac{R_2^2}{r^2}-1\right)\;,
$$
\be T^{\phi \phi}_{\rm walls}=-P
\frac{R_1^2}{R_2^2-R_1^2}\left(\frac{R_2^2}{r^2}+1\right)\;.\label{trans}\ee
Besides these transverse stresses, the magnetic field determines
also   axial stresses $T^{zz}_{\rm walls}(r)$ inside the walls. We
derive below the average value $F_{\rm ax}$ of $T^{zz}_{\rm
walls}$, as given in Eq. (\ref{compr}). In view of the key role
played by the axial compression $F_{\rm ax}$, and in order to
explain its physical origin, we provide two different derivations
of Eq. (\ref{compr}).
 The first
derivation is based on the general equilibrium conditions Eq.
(\ref{equil}). Indeed, using Eq. (\ref{equil}), one can prove the
following identity holding   far from  the solenoid's ends \be
\int_0^{R_2} dr\,2 \pi r\, \sum_{j=x,y}(T^{jj}_{\rm
walls}+T^{jj}_{\rm mag})
 {=}0\;.\label{corr}\ee   To obtain  it, we observe  that far
 from the end points, stresses are independent of $z$,
and therefore   Eqs. (\ref{equil}) reduce to: \be
\sum_{k=x,y}\partial_k \, (T^{jk}_{\rm walls}+
  \,T^{jk}_{\rm mag}) =0\,. \label{equilsol}\ee
Therefore, we have the identity: \be \sum_{j=x,y}( T^{jj}_{\rm
sol}+T^{jj}_{\rm mag})=\sum_{j,k=x,y}\partial_k\,[( T^{jk}_{\rm
sol}+T^{jk}_{\rm mag})\, x^j]\;.\ee Upon integrating both sides of
the above equation on a cross section $\Sigma$ of the solenoid, we
obtain:  $$ \int_0^{R_2} dr\,2 \pi r\, \sum_{j=x,y}(T^{jj}_{\rm
walls}+T^{jj}_{\rm mag})
 {=}$$
 \be= R_2 \int_0^{2 \pi} d\theta \sum_{j,k=x,y}(T^{jk}_{\rm walls}+T^{jk}_{\rm mag}) \,x^j\, n^k \vert_{r=R_2} \;.
 \label{intint}\ee
 The integral on the r.h.s. vanishes, because   $T^{jk}_{\rm
 walls} n^k|_{r=R_2}$ is zero in view of  Eq. (\ref{bc}),
 while $T^{jk}_{\rm
 mag} n^k|_{r=R_2}$ vanishes because the magnetic field is negligible outside a long
 solenoid (see the discussion of the external field following Eq. (\ref{magen})).
 Therefore, the l.h.s. of Eq. (\ref{intint}) is zero and this proves Eq.
 (\ref{corr}). Indeed, it
is easy to verify   that Eq.(\ref{corr}) is satisfied by   the
explicit expressions for the transverse material stresses given in
Eqs. (\ref{trans}), together with the Maxwell stresses Eq.
(\ref{Tijmag1}).

By using Eq. (\ref{corr}), we can now easily obtain $F_{\rm ax}$.
To do this, we recall the identity:
 \be
\int_0^{R_2} dr\,2 \pi r\, (T^{ii}_{\rm walls}+T^{ii}_{\rm mag})
 {=}\,0\;,\label{corrbis}\ee which is a direct consequence of Eq. (\ref{sigstr}).
 Upon subtracting Eq. (\ref{corr}) from
 Eq. (\ref{corrbis}), we then obtain:
 $$  \int_0^{R_2} dr\,2 \pi r\,(T^{zz}_{\rm walls}+T^{zz}_{\rm
 mag})=0\;.$$
It follows from the above Equation that \be F_{\rm ax}  =
\int_{R_1}^{R_2} dr\,2 \pi r\,T^{zz}_{\rm walls}= - \int_0^{R_2}
\!\!\!dr\,2 \pi r\,T^{zz}_{\rm mag}\;.\label{Fax2}\ee Upon using
into the r.h.s of the above formula the expression of $T^{zz}_{\rm
mag}$ inside the solenoid: \be T^{zz}_{\rm mag}= -\frac{B_{\rm
in}^2}{8 \pi}=-{\cal E}_{\rm mag}\,,\ee  we immediately obtain Eq.
(\ref{compr}).

In order to clarify the physical origin of the axial force $F_{\rm
ax}$, it is useful to provide a more direct derivation of Eq.
(\ref{compr}).
For this purpose, we consider the cylindrical  sheet ${\Sigma}$ of
radius $R_1$ and height $L$ that contains all the current flowing
in the solenoid, and we imagine splitting  it in two parts
${\Sigma}_1$ and $\Sigma_2$, consisting respectively of the points
of $\Sigma$ that lie above and below a plane of equation $z={\bar
z}$.
 If
we imagine $\Sigma_1$ and $\Sigma_2$ as consisting of a large
number of closed circular current loops, it is clear by Ampere's
law that an attractive axial force ${\bf F}^{(\rm Amp)}(\bar z)$
arises between $\Sigma_1$ and $\Sigma_2$, and we show below that
for $\bar z$ far from the end points ${\bf F}^{(\rm Amp)}(\bar z)$
has a constant magnitude equal to ${\tilde {\cal E}}_{\rm mag}$.

Indeed, axial symmetry implies that ${\bf F}^{(\rm Amp)}(\bar z)$
is along the $z$ axis, and we let $F_z^{(\rm el)}(\bar z)$ its
$z$-component.  Now, Ampere's law gives the following expression
for the elementary force $d{ F}_z^{(\rm Amp)}(z_1,z_2)$ between
two infinitesimal circular current loops  within $\Sigma_1$ and
$\Sigma_2$: \be d{F}_z^{(\rm Amp)}(z_1,z_2)=
-\frac{dj_1\,dj_2}{c^2} \oint \oint (\vec{dl}_1 \cdot \vec{dl}_2)
\frac{\vec{x}_{1}-\vec{x}_2}{|\vec{x}_1-\vec{x}_2|^3}\;,\ee where
$dj_i=j \,dz_i$, and ${\vec dl}_i$ are line elements tangential to
the surface elements, and parallel to the surface  current density
$\vec{j}$. Using cylindrical coordinates, the above integral can
be rewritten as:  \be d{ F}_z^{(\rm Amp)}  = - \frac{2 \pi\, j^2
R_1^2}{c^2}\, dz_1\,dz_2\int_{0}^{2 \pi} \!\!\!\!d\theta \frac{z
\cos \theta }{\sqrt{z^2+2 R_1^2(1-\cos \theta)}}\;,\ee with
$z=z_1-z_2$. Since the integrand is positive, we see that the two
rings attract each other, as expected. Upon integrating over $z_1$
and $z_2$ we then obtain for $ {F}_z^{(\rm Amp)}(\bar z)$ the
expression
 \be {F}_z^ {(\rm Amp)}(\bar z)= -  \frac{2 \pi^2 R_1^2
j^2}{c^2}\,I(\bar{z})\,,\label{force}\ee where $I(\bar{z})$ is the
integral \be I(\bar{z})=\frac{1}{\pi}\int_{\bar z}^{L/2}\!\!\!\!
dz_1 \int_{-L/2}^{{\bar z}} \!\!\!\!dz_2\int_0^{2 \pi} \!\!\!\!d
\theta\, \frac{z \cos \theta}{\sqrt{z^2+2 R_1^2(1-\cos
\theta)}}\;.\ee   The integrals over $z_1$ and $z_2$ in
$I(\bar{z})$ can be done by the change of variables $(z_1,z_2)
\rightarrow (w,z)$, where $w=(z_1+z_2)/2$. The result is:
\begin{widetext}$$ I(\bar z)= \frac{1}{\pi} \int_0^{2 \pi}\!\!\!\!d\theta
\left\{\log\left[1-\frac{2 \bar z}{L}+ \sqrt{ \left(1-\frac{2 \bar
z}{L}\right)^2\!\!\!+ 8\,\frac{R_1^2}{L^2}(1-\cos \theta)}\right]
+\log \left[1+\frac{2 \bar z}{L}+ \sqrt{ \left(1+\frac{2 \bar
z}{L}\right)^2\!\!\!+ 8\,\frac{R_1^2}{L^2}(1-\cos \theta)}
\right]+\right.$$ \be \left. -\log\left[1+ \sqrt{1+
2\,\frac{R_1^2}{L^2}(1-\cos \theta)}\right]-\frac{1}{2}\log(1-\cos
\theta) \right\} \cos \theta\,\ee\end{widetext}
where we omitted a few terms that are zero upon integrating over
$\theta$. The positive quantity $I(\bar z)$ reaches its maximum
value at the center of solenoid (for $\bar z=0$), and
monotonically decreases towards zero when $\bar z$ approaches the
end-points at $\pm L/2$. For a long solenoid, $R_1/L \ll 1$, and
far from the end points, $(L/2-|z|)/R_1 \gg 1$, $I(\bar z)$
becomes independent of $\bar z$ and its limiting value for an
infinitely long solenoid can be obtained by observing that for
$R_1/L \rightarrow 0$ the first three terms between the curly
brackets of the above integral become independent of $\theta$ and
therefore, after multiplication by $\cos \theta$, they integrate
to zero, leaving us with \be \lim_{
 R_1/L \rightarrow 0}
 I(\bar z)= - \int_0^{2 \pi}\!\!\frac{d\theta}{2 \pi}\,  \cos \theta \log(1-\cos
\theta) \,=1\,.\ee Upon inserting this value into Eq.
(\ref{force}), we see, as expected, that in the limit of a long
solenoid, and for $\bar{z}$  far from the end-points, the
  current sheets  $\Sigma_1$ and $\Sigma_2$ attract each other
with a force of magnitude   \be \lim_{ R_1/L \rightarrow 0}
F^{(\rm Amp)}(\bar z)= \frac{2 \pi^2 R_1^2
j^2}{c^2}\,.\label{inffor}\ee This  formula can be conveniently
expressed in terms of the magnetic energy density by noticing that
inside a long solenoid, the strength of the magnetic field $B_{\rm
in}$ is related to the surface current density as \be B_{\rm in}=
\frac{4 \pi\,j}{c}\;.\ee Using this formula, we can rewrite Eq.
(\ref{inffor}) as: \be  \lim_{ R_1/L \rightarrow 0}  F^{(\rm
Amp)}(\bar z)=\pi\, R_1^2\,\times \frac{B_{\rm in}^2}{8 \pi}
={\tilde {\cal E}}_{\rm mag}\;.\label{famplong}\ee
Obviously, since the current sheet $\Sigma$ is anchored to the
inner face of the solenoid, the electrodynamic force ${F}_z^{(\rm
Amp)}(\bar z)$ compresses the pipe and, at mechanical equilibrium,
it is balanced by the axial stresses $T^{zz}_{\rm walls}$ inside
the pipe walls: \be F_{\rm ax}(\bar z)={F}^{(\rm Amp)}(\bar
z)\;.\ee It should be noted that according to this formula the
compression $F_{\rm ax}(\bar z)$ vanishes at the ends of the pipe,
and it increases as one moves towards the middle of the pipe.  For
a very long pipe, Eq. (\ref{famplong}) implies that far from the
ends $F_{\rm ax}$ approaches the constant value: \be F_{\rm ax}=
{\tilde {\cal E}}_{\rm mag}\,,\ee which reproduces again Eq.
(\ref{compr}).

\end{document}